\documentclass{JHEP3}
\usepackage{amsmath}
\usepackage{mathrsfs}
\preprint{USTC-08-15\\ZIMP-08-21}

\title{On Moduli Stabilization Scheme in Type IIB Flux Compactifications}
\author{Huan-Xiong Yang \\ Interdisciplinary Center for Theoretical Study, \\
University of Science and Technology of China, \\
Hefei, 200026, P. R. China \\
\\Zhejiang Institute of Modern Physics,
\\ Physics Department, Zhejiang University, \\ Hangzhou, 310027, P.
R. China \\E-mail: \email{hxyang@zimp.zju.edu.cn}}

\date{\today}

\abstract{We revisit the two-stage procedure for moduli
stabilization in Type IIB orientifolds at light K\"ahler-modulus
limit. In view of the necessity to keep the K\"ahler geometry
structure of the moduli space during the stabilization, we define a
holomorphic quantity called effective superpotential. The KKLT
superpotential as well as the superpotential proposed by Villadoro
and Zwirner are then examined with respect to this holomorphic
effective superpotential. The mechanism is also illustrated with a
simple toy model of one complex structure modulus.
}


\begin{document}

Searching for de Sitter or Minkowski vacua in 4-dimensional spacetime with
spontaneously broken $\mathscr{N}=1$ supersymmetry is undoubtedly a challenge in
superstring phenomenology, which is directly related to the problem of moduli
stabilization. Recently, some promising progress has been made in this
direction, particularly in understanding moduli stabilization of Type IIB flux
compactifications on Calabi-Yau orientifolds. The first encouraging proposal
along this line was made by Kachru et al.\cite{hep-th/0301240}, now known as
KKLT in the literature. In the context of type IIB theory with D-branes, this
mechanism can be used for stabilizing all closed string moduli and constructing
de Sitter vacua through the incorporation of miscellaneous contributions of the
closed string background fluxes, the D-brane related nonperturbative effects and
the possible D-brane interactions. The idea of moduli stabilization is realized
by a two-stage decoupled procedure. In the first stage, we have an incomplete
F-term potential which is obtained by turning suitable 3-form fluxes on some
3-cycles of the Calabi-Yau orientifold and is independent of the K\"ahler-moduli
of the compactification, in particular the overall volume modulus. This
potential is then used to stabilize the complex structure moduli (including
dilaton-axion field) at their extrema at a high scale while the remaining light
K\"ahler moduli are kept free, therefore a low-energy no-scale effective
$\mathscr{N}=1$ supergravity theory with a constant superpotential $W_0$ is
formulated for the remaining light K\"ahler moduli. In the second stage, the
possible non-perturbative contributions from gauge theory on D7-branes and/or
from Euclidean D3-instantons, that can produce an exponential dependence on the
K\"ahler moduli in the superpotential, are included. These are in turn used to
stabilize the light K\"ahler moduli at their extrema. So long as $W_0 \ne 0$ and
the pullbacks of threefold Calabi-Yau divisors on a fourfold $X$ have arithmetic
genus {\em one}\cite {hep-th/0404257}, the stabilization of these light K\"ahler
moduli is possible. The resulting vacuum happens to be purely supersymmetric and
anti-de Sitter. KKLT further suggested to add the effect of anti D3-branes to
the model for the purpose of lifting the SUSY preserving AdS vacuum to a SUSY
breaking de Sitter vacuum.

KKLT proposal has raised a great enthusiasm in the research of
string moduli stabilization and de Sitter vacua recently. However
there still exist a few issues regarding the logical validity of
this proposal and its successful implementation within a realistic
string model. The {\em {ad hoc}} uplifting of the vacuum energy has
been questioned since the supersymmetry breaking introduced by an
anti-D3 brane is explicit and by hand. So far no consistent
supergravity description of such an approach has been found. There
are a few attempts trying to resolve this issue. The first
unsuccessful effort was by considering a D-term uplifting in
Ref.\cite{hep-th/0309187} but it turned out not to work since the
D-term considered there actually vanishes due to the vanishing F-term.
However, this is remedied in a recent paper \cite{hep-th/0601190}
by Ach\'ucarro et al (ADCD) where an anomalous D-term can be
generated due to an anomalous U(1). Other efforts for the uplifting
have been considered in Ref.\cite{hep-th/0508167} and in
Ref.\cite{hep-th/0602253}. In either case, the crucial thing is the
gauged U(1) symmetry which gives rise to a consistent D-term
responsible for the uplifting.

In almost all the models discussed above, the described
two-stage procedure of KKLT has been used for the moduli
stabilization. The validity of this two-stage procedure has been
raised recently in \cite{hep-th/0506266, hep-th/0506090,
hep-th/0511030} in which it was argued that the nonperturbative
contributions to the superpotential, if they exist at all, should be
considered logically throughout the whole stabilization process.
The KKLT two-stage procedure, if reliable,
must be a genuine light-K\"ahler-modulus approximation of such an
one-stage procedure. Including the possible nonperturbative effects
from the outset does generically violate the imaginary self-duality
(ISD) of the 3-form fluxes\cite{hep-th/0506090} (which is the
necessary condition for preserving $\mathscr{N}=1$ supersymmetry
during the stabilization of complex structure moduli and dilaton by
turning suitable 3-form fluxes\cite{hep-th/0105097}) so  the
supersymmetry is spontaneously broken at the expected F-term
potential level. In general, the one-stage procedure is too
complicated and it is hard to deal with. Fortunately, in
Ref.\cite{hep-th/0506266} de Alwis proposed a modified version of
the two-stage procedure, in which the imaginary self-duality of the
3-form fluxes is persisted and then the supersymmetry should not be
broken during the moduli stabilization (if we ignore the
contributions of massless squark condensation\cite{hep-th/0601190},
for example). The difference between KKLT original procedure and de
Alwis' improvement is that in de Alwis' approach, the
nonperturbative corrections to the superpotential are considered at
both stages for moduli stabilization. de Alwis examined the KKLT
procedure within his modified prescription and concluded that the
original decoupled two-stage procedure can {\em not} be viewed as a
proper approximation (at the light-K\"ahler-modulus limit) to the
exact one-stage procedure.

Given that many models for uplifting (to the metastable deSitter vacua) are
based on the KKLT two-stage moduli stabilization procedure, the justification of
this decoupled procedure within the strict one-stage approach seems really
imperative. In this short note we are going to make such a justification. As in
Ref.\cite{hep-th/0506266} by de Alwis, we take light K\"ahler-modulus
approximation (light-$T$ limit for short from now on), insist on the imaginary
self-duality of 3-form fluxes during the moduli stabilization and pay main
attention to  K\"ahler function $G(T,~\bar{T})= -3\ln(T+\bar{T}) +\ln\Lambda
(Ce^{- h T}, \bar{C}e^{- h \bar{T}})$\footnotemark[1]. Here $\Lambda (Ce^{- h
T}, \bar{C} e^{- h \bar{T}})$ is defined as a power series expansion in terms of
nonperturbative superpotential $Ce^{-h T}$ (See below). The new ingredient in
our approach is that we put forward a concept of {\em effective superpotential}.
This effective superpotential with holomorphicity is necessary if the two-stage
moduli stabilization procedure at the light-$T$ limit preserves the K\"ahler
geometry structure of the moduli space. The mathematical structure of the KKLT
superpotential, {\em a constant plus a term proportional to the nonperturbative
superpotential}, can be interpreted as this effective superpotential if we keep
only the leading order terms in $\Lambda (Ce^{- h T}, \bar{C}e^{- h \bar{T}})$.
Moreover, we observe that the Villadoro-Zwirner
superpotential\cite{hep-th/0508167}, which consists of a single $T$-dependent
exponential and was regarded to be very difficult to have a stringy
realization\cite{hep-th/0602253, hep-th/0602120}, will be realized as an
effective superpotential if the power series $\Lambda (Ce^{- h T}, \bar{C}e^{- h
\bar{T}})$ is approximated up to the second order terms $\mathscr{O}(C^2e^{-2 h
T})$. \footnotetext[1]{In $\mathscr{N}=1$ supergravity, $G(T, \bar{T})$ is
defined as $G(T, \bar{T})=K(T, \bar{T}) + \ln |W(T)|^2$, which is invariant
under the K\"ahler transformation $K(T, \bar{T})\rightarrow K(T, \bar{T}) + \ln
|f(T)|^2, ~~W(T) \rightarrow \dfrac{1}{f(T)}W(T)$.}

We now begin to report our results in detail. Firstly, let us give a
brief review of the modified KKLT mechanism \cite{hep-th/0506266}
and the improved uplifting prescription proposed in
Ref.\cite{hep-th/0601190}.  As was shown in
Ref.\cite{hep-th/0403067}, the low energy gauge-invariant action for
$\mathscr{N}=1, D=4$ supergravity with chiral and vector multiplets
coming from Type IIB string theory compactified on a Calabi-Yau
orientifold is determined by four ingredients: the real K\"ahler
potential $K$, the holomorphic superpotential $W$, the holomorphic
gauge kinetic function $f_{ab}$ and the holomorphic Killing vectors.
It is remarkable that although the D-term part of the supergravity
scalar potential is governed by all these ingredients, the F-term
potential is solely determined by the real K\"ahler potential $K$
and the holomorphic superpotential $W$,
\begin{equation}\label{eq: 1}
V_F = e^{K}(K^{i\bar{j}}D_{i}W D_{\bar{j}}\bar{W} -3|W|^2)
\end{equation}
or equivalently determined by the invariant K\"ahler function $G$,
\begin{equation}\label{eq: 2}
V_F = e^{G}(G^{i\bar{j}}G_{i}G_{\bar{j}}-3)
\end{equation}
where $G=K+\ln|W|^2$. For simplicity, here we only consider the
orientifolds with just one overall K\"ahler modulus $T$. The
classical K\"ahler potential\footnotemark[2] and the superpotential
coming from fluxes and the nonperturbative contributions of the
Euclidean instanton or gaugino condensation are then,
\footnotetext[2]{For simplicity here we only consider the tree-level
K\"ahler potential. However, inclusion of the possible perturbative
and nonperturbative corrections to K\"ahler potential into the present
scenario is straightforward.}
\begin{equation}\label{eq: 3}
\left.
\begin{array}{lll}
 K & = & -3 \ln ( T+\bar{T})-\ln (S+\bar{S} )+ N_f |M|^2 + k( z^i, {\bar{z}}^{\bar{j}} ) \\
 W & = & A( z^i ) + S B( z^i ) +C e^{-h T}
\end{array}
\right.
\end{equation}
where $C=(N-N_f)(\frac{2}{M^{2N_f}})^{\frac{1}{N-N_f}}$ and
$h=-\frac{2N_f(q+\bar{q})}{\delta_{GS}(N-N_f)}$. Here the
nonperturbative contribution to the superpotential is from the
condensation $(M^2)^{i}_{j} =2Q^i \bar{Q}_j$ of $N_f$ massless
squark pairs $\{ Q^i, \bar{Q}_j \}$ with color group
$U(N)=SU(N)\times U_X(1)$\cite{hep-th/0601190}. $\delta_{GS}$, $q$
and $\bar{q}$ are the charges of K\"ahler modulus $T$ and the squark
and anti squark under the anomalous gauge group $U_X(1)$,
respectively. Under a $U_X(1)$ transformation the K\"ahler modulus
transforms as $T \rightarrow T+i \frac{\delta_{GS}}{2} \epsilon$
while $Q^i \rightarrow e^{i q \epsilon} Q^i$ and $\bar{Q}_i
\rightarrow e^{i \bar{q} \epsilon}\bar{Q}_i$. The K\"ahler potential
is manifestly invariant under $U_X(1)$ transformation, and the
anomaly cancelation condition\cite{hep-th/0601190} further
guarantees the invariance of the superpotential. Provided that the
superpotential is reduced to the KKLT type $W=W_0 + Ce^{-h T}$ with
$W_0$ an effective constant (after the complex structure moduli and
dilaton-axion are integrated out) in the light-$T$ limit, the
authors of Ref.\cite{hep-th/0601190} showed that the unbroken
supersymmetry conditions $D_T W = D_M W =0$ cannot be simultaneously
fulfilled for the F-term potential $V_F$, therefore indicating that
the minimum of $V_F$ is at a supersymmetry breaking point in moduli
space. A nonvanishing D-term potential can then be added to uplift
the vacuum to be de Sitter \cite{hep-th/0309187, hep-th/0108200,
hep-th/0402046}.

Now the question is whether the superpotential can be cast as $W=W_0 + Ce^{-h
T}$ in general in the light-$T$ limit. In the original KKLT scheme, such a
superpotential was obtained via a decoupled two-stage procedure. KKLT first
considered the 3-form fluxes $G_3 =F_3 + iS H_3$ to give rise to the
superpotential $W_{flux}:= \dfrac{1}{(2\pi)^2 \alpha^{\prime}}\int \Omega \wedge
G_3 =A( z^i ) + S B( z^i )$  \cite{hep-th/9906070} and used it to fix the
complex structure moduli and dilaton-axion field via the supersymmetric extreme
conditions $D_{z^i} W_{flux} = D_S W_{flux}=0$. The validity of these conditions
indicates that in the process of fixing the complex structure moduli (including
dilaton-axion) the imaginary self-duality $G_3=-i*_6G_3$ of 3-form fluxes is
preserved\cite{hep-th/0105097}. After that, the nonperturbative corrections to
superpotential was added to define the ``full" superpotential $W=W_0 + Ce^{-h
T}$, with $W_0$ the value of $W_{flux}$ at the supersymmetric minimum of F-term
potential energy. The reliability of this {\em ad hoc} two-stage procedure was
assumed in Ref.\cite{hep-th/0301240} on the argument that in some region of
moduli space the complex structure moduli and dilaton-axion field are heavy
enough so they could be integrated out with the partial superpotential purely
from the flux contributions. However, de Alwis pointed out that such an argument
is untrustworthy if one takes the light-$T$ approximation
seriously\cite{hep-th/0506266}. The point in de Alwis' analysis is that the full
superpotential in Eqs.(\ref{eq: 3}) should be used to solve the vanishing F-term
conditions
\begin{equation}\label{eq: 4}
\left.
\begin{array}{l}
D_S W =  B( z^i ) -(S +\bar{S} )^{-1} W =0 \\
D_j W = \partial_j A(z^i) + S \partial_j B(z^i) + \partial_j k(z^i, \bar{z}^i) W =0
\end{array}
\right.
\end{equation}
The non-perturbative term $Ce^{-h T}$ considered explicitly in the full
superpotential means that the dilaton-axion field $S$ and the complex structure
moduli $z^i ~[i=1,~2,~\cdots,~h^{(2,1)}]$ can not be integrated out
holomorphically through the solutions of Eqs.(\ref{eq: 4}), which in general
depend on both of $Ce^{-hT}$ and its conjugate. Technically, due to the highly
nonlinearity of Eqs.(\ref{eq: 4}), exact solutions are in general not expected.
Fortunately, at the light-$T$ limit for which $h \Re{T} \gg 1 $, approximated
solutions can be found through a power series expansion as follows,
\begin{equation}\label{eq: 5}
\left.
\begin{array}{lll}
S & = & \alpha + \beta C e^{- h T} + \gamma \bar{C} e^{- h \bar{T}}
+ \zeta C^2 e^{-2 h T} + \chi {\bar{C}}^2 e^{-2 h \bar{T}}
+ \psi |C|^2 e^{-h (T + \bar{T})} + \cdots \\
z^i & = & {\alpha}^i + {\beta}^i C e^{- h T} + {\gamma}^i \bar{C} e^{- h \bar{T}}
+ {\zeta}^i C^2 e^{-2 h T}
+ {\chi}^i {\bar{C}}^2 e^{-2 h \bar{T}} + {\psi}^i |C|^2 e^{-h (T + \bar{T})}
+ \cdots
\end{array}
\right.
\end{equation}
In either one-stage procedure\cite{hep-th/0506090, hep-th/0511030} or the
modified two-stage procedure of moduli stabilization developed by de
Alwis\cite{hep-th/0506266}, the heavy complex structure moduli and the
dilaton-axion field are integrated out through Eqs.(\ref{eq: 5}). Different from
the original KKLT two-stage procedure, the non-K\"ahler moduli have not been
fixed at their extreme values hereunto, instead they are integrated out as
functions of both $Ce^{-hT}$ and $\bar{C}e^{-h\bar{T}}$. The resultant
superpotential does generically become a nonholomorphic function of the K\"ahler
modulus $T$, {\em i.e.}, $W=W(Ce^{-h T}, \bar{C} e^{-h \bar{T}})$. In the
one-stage procedure, the F-term potential $V_F$ is found to be
\begin{equation}\label{6a}
\begin{array}{lll}
    V_F & = & e^K \big[ \frac{1}{3} (T + \bar{T})^2 |D_T W|^2 - 3|W|^2 \big ]\\
        & = & \frac{e^{N_f|M|^2}}{(T + \bar{T})^2} \cdot \frac{1}{(S +\bar{S})}
e^{k(z^i,~\bar{z}^{\bar{j}})} \bigg [ \frac{1}{3}(T + \bar{T}) h^2 |C|^2 e^{-h
(T + \bar{T})} + h (W \bar{C} e^{-h \bar{T}} + \bar{W} C e^{-h T}) \bigg ]
\end{array}
\end{equation}
where the K\"ahler derivative of superpotential with respect to the K\"ahler
modulus $T$
\[
D_T W = -h Ce^{-hT} -\frac{3W}{(T + \bar{T})}
\]
has been used. By Eqs.(\ref{eq: 5}) the superpotential is expressed as,
\begin{equation}\label{eq: 7}
W = \xi_0 + \xi_{10} C e^{-h T} + \xi_{01} \bar{C} e^{- h \bar{T}} + \xi_{20}
C^2 e^{-2 h T} + \xi_{02} \bar{C}^2 e^{-2 h \bar{T}} + \xi_{11} |C|^2 e^{-h (T +
\bar{T})} + \cdots
\end{equation}
Similarly, the invariant K\"ahler function that depends only upon modulus $T$
and its conjugate after the non-K\"ahler moduli are integrated out
becomes\cite{hep-th/0506266} \footnotemark[3], \footnotetext[3]{The $U_X (1)$
symmetry still remains.}
\begin{equation}\label{eq: 8}
\left.
\begin{array}{ll}
G = & -3 \ln (T + \bar{T}) + N_f |M|^2  \\
& + \ln \Big [ v + b Ce^{-h T} + \bar{b} \bar{C} e^{-h \bar{T}} + c C^2 e^{-2 h
T} + \bar{c} \bar{C}^2 e^{-2 h \bar{T}} + d |C|^2 e^{- h (T + \bar{T})} + \cdots
\Big ]
\end{array}
\right.
\end{equation}
Based on these two series solutions to $W$ and $G$, we can formulate the F-term
potential energy as:
\begin{equation}\label{eq: 9}
\begin{array}{lll}
V_F & = & \frac{e^{N_f|M|^2}}{|\xi_0|^2 (T + \bar{T})^2} \bigg [ h v \bar{\xi}_0
Ce^{-h T} + h v \xi_0 \bar{C}e^{-h \bar{T}}  + hv \bar{\xi}_{01} C^2 e^{-2 h T}
+ h b \bar{\xi}_{0} C^2 e^{-2 h T} \\
&  & ~~ ~~ - \frac{h v (\xi_0 \bar{\xi}_{01}+ \xi_{10}\bar{\xi}_0)}{\xi_0} C^2
e^{-2 h T}
+ hv \xi_{01} \bar{C}^2 e^{-2 h \bar{T}}  + h \bar{b} \xi_0 \bar{C}^2 e^{-2 h \bar{T}}\\
&  & ~~ ~~ - \frac{h v (\xi_0\bar{\xi}_{10} + \xi_{01} \bar{\xi}_0
)}{\bar{\xi}_0} \bar{C}^2 e^{-2 h \bar{T}}
+ \frac{1}{3}(T + \bar{T}) h^2 v |C|^2 e^{-h (T + \bar{T})} \\
&  & ~~ ~~ + hv(\xi_{10}+ \bar{\xi}_{10})|C|^2 e^{-h (T + \bar{T})}
+ h (b\xi_0 + \bar{b} \bar{\xi}_0) |C|^2 e^{-h (T + \bar{T})}  \\
&  & ~~ ~~ - \frac{h v (\xi_0\bar{\xi}_{01} + \xi_{10} \bar{\xi}_0
)}{\bar{\xi}_0}|C|^2
e^{-h (T + \bar{T})}\\
&  & ~~ ~~ -\frac{h v (\xi_0\bar{\xi}_{10} + \xi_{01} \bar{\xi}_0 )}{\xi_0}
|C|^2 e^{-h (T + \bar{T})}
+ \mathscr{O}(C^3e^{-3hT}) \bigg ]\\
\end{array}
\end{equation}
Recall that the supersymmetric extremes are determined by conditions $D_SW =
D_jW = D_T W=0$. These vacua (if exist) are bounded to be anti-deSitter spaces
with potential energy $V_F = -3 e^K |W|^2<0$. At such a supersymmetric vacuum
the K\"ahler modulus $T$ is frozen by the solutions of the equation
$\frac{3W}{(T + \bar{T})} = hCe^{-hT}$ that becomes
\[
\begin{array}{ll}
& \frac{3}{(T + \bar{T})}\Big[\xi_0 + \xi_{10} C e^{-h T} + \xi_{01} \bar{C}
e^{- h \bar{T}}
+ \xi_{20} C^2 e^{-2 h T} + \xi_{02} \bar{C}^2 e^{-2 h \bar{T}} \\
& ~~~~~~~~~~~~~ + \xi_{11} |C|^2 e^{-h (T + \bar{T})} + \cdots \Big ] =
hCe^{-hT}
\end{array}
\]
in the light-$T$ limit. Of course, there are probably some non-supersymmetric
vacua determined by conditions $D_SW = D_jW = \partial_T V_F = 0$
\footnotemark[4]. \footnotetext[4]{The general supersymmetry breaking vacua are
determined by $\partial_S V_F = \partial_j V_F = \partial_T V_F =0$. However, we
are interested in those vacua in which the supersymmetries are minimally
broken.}

Despite reliable in principle, the one-stage procedure is generically too fussy
in technique to be used in practical moduli stabilization. The two-stage
procedure suggested by KKLT \cite{hep-th/0301240}, on the other hand, is much
simpler. From the perspective of two-stage procedure, however, the
nonholomorphicity of the superpotential (as the function of light K\"ahler
modulus) in the second stage is absolutely unacceptable. Non-holomorphicity of
the superpotential implies the spoilage of the K\"ahler geometry structure of
the moduli space (in view of the two-stage procedure).  Although the invariant
function $G$ itself is allowed to be nonholomorphic, the equivalence between the
two expressions of the F-term potential, $V_F=e^G(G^{T\bar{T}}G_T
G_{\bar{T}}-3)$ and $V_F = e^K (K^{T\bar{T}}|D_T W|^2 - 3 |W|^2)$, depends
crucially upon the holomorphicity of the superpotential $W$ appearing in the
relation $G=K+\ln|W|^2$. For this reason, in the two-stage procedure, it is
still preferable to have an {\em effective holomorphic superpotential}
$W_{\textrm{eff}}(T, M)$ so that
\begin{equation}\label{eq: 10}
G = -3 \ln (T + \bar{T}) + N_f |M|^2 + \ln |W_{\textrm{eff}}(T, M)|^2
\end{equation}
and then the usual practice applies:
\begin{equation}\label{eq: 11}
V_F = \frac{e^{N_f |M|^2}}{(T + \bar{T})^2} \bigg[\frac{1}{3}(T + \bar{T})
|\partial_T W_{\textrm{eff}}|^2 - \big ( W_{\textrm{eff}}\partial_{\bar{T}}
\bar{W}_{\textrm{eff}} + \bar{W}_{\textrm{eff}}\partial_T W_{\textrm{eff}} \big
) \bigg]
\end{equation}
The consistency of two expressions (\ref{eq: 9}) and (\ref{eq: 11}) for the
F-term potential energy $V_F$ up to the second order of $Ce^{-hT}$ implies that
the effective superpotential $W_{\textrm{eff}}(T, M)$, if it exists,  should
take the KKLT form,
\begin{equation}\label{eq: 12}
    W_{\textrm{eff}}(T, M) \approx   W_0 + g Ce^{-hT}
\end{equation}
$W_{\textrm{eff}}(T, M)$ is in principle determined by the equivalence
\begin{equation}\label{eq: 13}
|W_{\textrm{eff}}(T, M)|^2 = \Big [ v + b Ce^{-h T} + \bar{b} \bar{C} e^{-h
\bar{T}} + c C^2 e^{-2 h T} + \bar{c} \bar{C}^2 e^{-2 h \bar{T}} + d |C|^2 e^{-
h (T + \bar{T})} + \cdots \Big ]
\end{equation}
between the two expressions (\ref{eq: 8}) and (\ref{eq: 10}) of the invariant
K\"ahler function $G=G(T, \bar{T})$. The verification of these equivalences can
actually be fulfilled order by order with respect to $Ce^{-h T}$ (and its
complex conjugate) in light $T$ limit. To the first order of $Ce^{-h T}$, the
coincidence of the two expressions of the F-term potential energy calculated
within two different procedures demands $v=b\xi_0$, under which the effective
superpotential exists and is given by Eq.(\ref{eq: 12}) with $|W_0|^2=v$ and
$g\bar{W}_0=b$ (Here we suppose $v \ne 0$)\footnotemark[5]. To the second order
of $Ce^{-h T}$, the equivalence of Eqs.(\ref{eq: 9}) and (\ref{eq: 11}) imposes
more severe constraints $b \xi_0 =d \xi_0^2 =v,~ 2c\xi_0 = b (1-\xi_{10}) $ and
$\xi_{01}=0$ on the parameters of the series solutions (\ref{eq: 7}) and
(\ref{eq: 8}). Eq.(\ref{eq: 13}) is also very stringent that has no solutions
unless $b^2 =vd$\footnotemark[6]. In particular, if the parameters
$v=b=c=0,~\xi_0=0, ~\xi_{10}=1, ~\xi_{01}=0$ but $d>0$, the two procedures are
equivalent up to $\mathscr{O}(C^2e^{-2hT})$, with the effective superpotential
in the two-stage procedure given by
\begin{equation}\label{eq: 14}
W_{\textrm{eff}}(T, M) = \sqrt{d} Ce^{- h T}
\end{equation}
It has the very form suggested in Ref.\cite{hep-th/0508167} for having a
nonvanishing D-term uplift, thought difficult there in having an explicit
stringy realization\cite{hep-th/0602120, hep-th/0602253}\footnotetext[5]{If $v =
0$, the effective superpotential of the KKLT type does not exist.}.
\footnotetext[6]{The condition $b^2=vd$ is generally an independent constraint
although it has been implied by $b \xi_0 =d \xi_0^2 =v$ when $\xi_0 \ne 0$.}

We are now at the position to give a toy model to illustrate how the effective
superpotential works. Consider a Type IIB orientifold with one overall K\"ahler
modulus and just one complex structure modulus. The tree-level K\"ahler
potential and the superpotential (from $3$-from flux contribution plus
nonperturbative corrections) read\cite{hep-th/0506090, hep-th/0511030},
\begin{equation}\label{eq: 15}
\left.
\begin{array}{lll}
K & = & -3\ln (T + \bar{T}) - \ln (S + \bar{S}) - \ln (U + \bar{U}) \\
W & = & \alpha_0 + \alpha_1 U + \alpha_2 S + \alpha_3 SU + Ce^{- h T}~~~~~
(\alpha_i \in \mathbf{R}, ~~h >0)
\end{array}
\right.
\end{equation}
For simplicity  we ignore the contributions from the open string moduli, {\em
i.e.}, a constant prefactor $C$ is assumed in the above
equations\footnotemark[7]. We also assume that all of the flux parameters
$\alpha_i$ ~$(i=0,\cdots,3)$ are non-vanishing. \footnotetext[7]{Though the
massless squark pair condensation $M$ is crucial for having a
supersymmetry-broken F-term vacuum\cite{hep-th/0601190}, it is not important
here in searching for the effective superpotential $W_{\textrm{eff}}(T, M)$.}

In terms of the full superpotential given in Eq.(\ref{eq: 15}), we can
explicitly write down the supersymmetry-preserving extreme conditions $D_U W
=D_S W =0$:
\begin{equation}\label{eq: 16}
\left.
\begin{array}{l}
\alpha_0 +\alpha_1 U - \alpha_2 \bar{S}  - \alpha_3 \bar{S}U + Ce^{-h T} =0 \\
\alpha_0 -\alpha_1 \bar{U} + \alpha_2 S  - \alpha_3 S\bar{U} + Ce^{-h T} =0 \\
\end{array}
\right.
\end{equation}
By setting $U=u+i\nu$, $S=s+i\sigma$, $X = \frac{1}{2}(C e^{-h T} + \bar{C} e^{-
h \bar{T}})$ and $Y = \frac{1}{2i}(C e^{-h T} - \bar{C} e^{- h \bar{T}})$, we
can recast Eqs.(\ref{eq: 16}) as,
\begin{equation}\label{eq: 17}
\left.
\begin{array}{l}
    \alpha_0 - \alpha_3 (su + \sigma \nu) +X =0 \\
    s\nu - \sigma u =0 \\
        \alpha_2 s - \alpha_1 u =0 \\
    \alpha_1 \nu + \alpha_2 \sigma + Y = 0
\end{array}
\right.
\end{equation}
From Eqs.(\ref{eq: 17}) we see that to ensure both $s$ and $u$ being fixed at
the acceptable positive values $\alpha_1$ and $\alpha_2$ should have the same
sign. Having a meaningful solution of this set of equations in the light-$T$
limit does also require both $\alpha_0$ and $\alpha_3$ have the same sign.
Eqs.(\ref{eq: 17}) are reduced to $\nu = -\frac{Y}{2\alpha_1}$,
$\sigma=-\frac{Y}{2\alpha_2}$, $s=\frac{\alpha_1u}{\alpha_2}$ and
\[
4\alpha^2_1 \alpha_3 u^2 -4 \alpha_0 \alpha_1 \alpha_2 - 4 \alpha_1 \alpha_2 X
+ \alpha_3 Y^2 =0
\]
In the light-$T$ limit, by solving the last equation we can write $u$ as a power
series expansion in $X$ and $Y$,
\begin{equation}\label{eq: 18}
u= \zeta \bigg [ 1 + \frac{X}{2 \alpha_0}  - \frac{X^2}{8 \alpha_0^2}
- \frac{\alpha_3 Y^2 }{8 \alpha_0 \alpha_1 \alpha_2 }  \bigg ] + \cdots
\end{equation}
with $\zeta =\sqrt{\frac{\alpha_0 \alpha_2}{\alpha_1 \alpha_3}}$ a
real and positive parameter. The remaining equations determine the rest and
up to the second order of $Ce^{-hT}$ the moduli $S$ and $U$ are expressed as:
\begin{equation}\label{eq: 19}
\left.
\begin{array}{lll}
S &  \approx  & \frac{\alpha_1 }{\alpha_2}\zeta \Big [ 1 + \frac{X}{2 \alpha_0}
- \frac{X^2}{8 \alpha_0^2}
- \frac{\alpha_3 Y^2 }{8 \alpha_0 \alpha_1 \alpha_2 }  \Big ]
-i \frac{Y}{2\alpha_2} \\
U &  \approx  & \zeta \Big [ 1 + \frac{X}{2 \alpha_0}
- \frac{X^2}{8 \alpha_0^2}
- \frac{\alpha_3 Y^2 }{8 \alpha_0 \alpha_1 \alpha_2 }  \Big ]
-i \frac{Y}{2\alpha_1}
\end{array}
\right.
\end{equation}
These equations are used to integrate out the complex structure modulus and
dilaton-axion field. After that, the superpotential acquires the form of
Eq.(\ref{eq: 7}) with the parameters given by,
\begin{equation}\label{eq: 20}
\begin{array}{l}
  \xi_0 = 2 (\alpha_0 + \alpha_1 \zeta)\\
  \xi_{10} = 1 + \frac{\zeta}{2\alpha_0 \alpha_2}(\alpha_1 \alpha_2 -\alpha_0
  \alpha_3)\\
  \xi_{01} = 1 + \frac{\zeta}{2\alpha_0 \alpha_2}(\alpha_1 \alpha_2 +\alpha_0
  \alpha_3)\\
  \xi_{20} = \frac{\alpha_3}{8\alpha_1 \alpha_2} - \frac{\zeta}{16 \alpha_0^2
  \alpha_2} (\alpha_1 \alpha_2 + \alpha_0 \alpha_3)\\
  \xi_{02} = \frac{\alpha_3}{8\alpha_1 \alpha_2} - \frac{\zeta}{16 \alpha_0^2
  \alpha_2} (\alpha_1 \alpha_2 - 3\alpha_0 \alpha_3)\\
  \xi_{11} = - \frac{\alpha_3}{4\alpha_1 \alpha_2} - \frac{\zeta}{8 \alpha_0^2
  \alpha_2} (\alpha_1 \alpha_2 + \alpha_0 \alpha_3)
\end{array}
\end{equation}
As expected, this naive superpotential depends only on the K\"ahler modulus $T$
but in a non-holomorphic manner. Plugging Eqs.(\ref{eq: 19}), (\ref{eq: 7}) and
(\ref{eq: 20}) into $G=K+\ln|W|^2$ we get the anticipated result
\begin{equation}\label{eq: 21}
\left.
\begin{array}{l}
G(T, \bar{T})  =  -3\ln (T + \bar{T}) + \ln \Lambda (Ce^{- h T}, \bar{C}e^{- h \bar{T}}) ,~\\
\Lambda (Ce^{- h T}, \bar{C}e^{- h \bar{T}})  =  \frac{|W|^2 }{(S + \bar{S}) (U + \bar{U})}  \\
~~~~~~~~~~~~~~~~~~~~~~~ \approx \Big [ v + b ( Ce^{-h T} + \bar{C} e^{-h \bar{T}}) + c ( C^2 e^{-2 h T} +
                          \bar{C}^2 e^{-2 h \bar{T}} ) + d |C|^2 e^{- h (T + \bar{T})} + \cdots \Big ]~,
\end{array}
\right.
\end{equation}
with parameters as follows,
\begin{equation}\label{eq: 22}
\left.
\begin{array}{lll}
v & = & \frac{\alpha_3  }{\alpha_0} (\alpha_0 + \alpha_1 \zeta )^2 \\
b & = & \frac{\alpha_3 }{2 \alpha_0}  (\alpha_0 + \alpha_1 \zeta ) \\
c & = & - \frac{ \alpha_3 \zeta} {16 \alpha_0^2 \alpha_2}
(\alpha_1 \alpha_2 - \alpha_0 \alpha_3 )\\
d & = & - \frac{ \alpha_3 \zeta}{8 \alpha_0^2 \alpha_2 } (\alpha_1 \alpha_2 + \alpha_0 \alpha_3)
\end{array}
\right.
\end{equation}
Therefore, for the model under consideration, the  expansion of
$\Lambda (Ce^{- h T}, \bar{C}e^{- h \bar{T}}) $  to the first order
of $Ce^{-h T}$ admits  a KKLT-like effective superpotential
\begin{equation}\label{eq: 23}
W_{\textrm{eff}}(T) = W_0 + g Ce^{- h T}
\end{equation}
through $\Lambda \approx |W_{\textrm{eff}}(T)|^2$, with $W_0 \approx
\sqrt{\frac{\alpha_3}{\alpha_0}}(\alpha_0 + \alpha_1 \zeta )$ and $g \approx
\frac{1}{2}\sqrt{\frac{\alpha_3}{\alpha_0}} $. The requirement $b\xi_0=v$ for
having a correct $V_F$ (up to the first order of $Ce^{-hT}$) in the two-stage
procedure is also in saturation. Notice if the original KKLT two-stage
procedure\cite{hep-th/0301240} is employed, the above two constants would be
given as  $W_0 = 2 (\alpha_0 + \alpha_1 \zeta )$ and $g=1$ instead. Because the
concept of effective superpotential  in the modified two-stage procedure has its
root in the exact one-stage procedure developed by L\"ust et
al\cite{hep-th/0506090}, the similar form between $W_{\textrm{eff}}(T)$ in
Eq.(\ref{eq: 23}) and the superpotential assumed in the original {\em
{decoupled}} KKLT procedure may indicate the validity of the original KKLT
procedure in sense of the effective superpotential $W_{\textrm{eff}}(T)$ in this
modified light-$T$ limit.

The existence of the effective superpotential at the ${\mathscr O}(C^2e^{-2 h
T})$ level of the $\Lambda$'s expansion is a genuine challenge because of the
obligatory constraints $v^2 = bd$ and $\xi_{01}=0$\footnotemark[8]. These
constraints turn out to be so stringent that they are not satisfied in general.
One interesting exception we find occurs if the flux parameters accidentally
satisfy $\zeta= - \alpha_0/\alpha_1$ in which $v=b=c=\xi_{01}=0$ and
$d=\frac{\alpha_3^2}{4\alpha_1 \alpha_2}>0$. Consequently, there exists an
effective superpotential of the exponential type\cite{hep-th/0508167},
\footnotetext[8]{The remaining conditions $b\xi_0=v$ and $2c\xi_0 =
b(1-\xi_{10})$ have been satisfied. See Eqs.(\ref{eq: 20}) and (\ref{eq: 22}).}
\begin{equation}\label{eq: 24}
W_{\textrm{eff}}(T) = \frac{\alpha_3}{2\sqrt{\alpha_1 \alpha_2}}
Ce^{- h T}.
\end{equation}
 This superpotential does not yield any minima in F-term potential
$V_F$, for which it was regarded to be unacceptable in the original
KKLT procedure. However, it provides an appealing mechanism to break
(spontaneously) supersymmetry. As mentioned earlier, such a
holomorphic superpotential has been used by Villadora and Zwirner to
establish a consistent D-term uplift scenario\cite{hep-th/0508167},
where both $V_F$ and $V_D$ are monotonic functions of the K\"ahler
modulus $T$ and their sum (the full potential) has a possibility of
giving a  de Sitter vacuum where the modulus $T$ is fixed.

In conclusion, we have revisited de Alwis' modified two-stage procedure at
light-$T$ limit and reexamined the KKLT superpotential within such a scenario.
This modified approach is found to be useful if the invariant K\"ahler function,
after the non-K\"ahler moduli are integrated out, can be expressed in terms of
an {\em effective superpotential} with holomorphicity. The invariant K\"ahler
function consists of two terms of which both are logarithmic functions of the
K\"ahler modulus. The first logarithmic function is nothing but the tree-level
K\"ahler potential for K\"ahler modulus. The second logarithmic function, on the
other hand, depends upon the K\"ahler modulus through a variable $\Lambda$ which
is generically a power series expansion of the nonperturbative superpotential
and its complex conjugate. The KKLT-like superpotential turns out to appear as
the leading-order approximation of the effective superpotential if $\Lambda$
contains a nonvanishing constant term, otherwise it can not be understood within
de Alwis's approach. Besides, Villadora-Zwirner superpotential may be produced
as an effective superpotential in this scenario if we approximate $\Lambda$ to
the second order of the nonperturbative superpotential. The analysis in the
context can be straightforwardly extended to the multi-K\"ahler moduli
orientifolds, from which the so-called ``better racetrack"
superpotentials\cite{hep-th/0404257, hep-th/0603129} are expected to be
reproduced as some appropriately defined effective superpotentials.

\section*{Acknowledgments}
I am deeply grateful to Prof. J.-X. Lu for his kind hospitality during my
visiting USTC and his valuable comments and fruitful suggestions on this
manuscript. This work is partially supported by CNSF-10375052, CNSF-10247009 and
the Startup Foundation of the Zhejiang Education Bureau. Acknowledged is also
the support through the USTC-ICTS by grants from the Chinese Academy of Sciences
and a grant from the NSF of China with 10535060.

\bibliography{StringCosmology}
\bibliographystyle{JHEP}
\end{document}